\documentclass[twocolumn,showpacs]{revtex4} %
\usepackage{epsfig,amsmath,amssymb,graphicx,bm}
\usepackage[english]{babel}

\begin{document}

\title{ MODIFIED PERTURBATION THEORY FOR THE YUKAWA MODEL} %
\author{Yu.M.\,Poluektov}
\email{yuripoluektov@kipt.kharkov.ua} %
\affiliation{%
National Science Center ``Kharkov Institute of Physics and
Technology'', 1, Akademicheskaya St., 61108 Kharkov, Ukraine }%

\begin{abstract}
A new formulation of perturbation theory for a description of the
Dirac and scalar fields (the Yukawa model) is suggested. As the main
approximation the self-consistent field model is chosen, which
allows in a certain degree to account for  the effects caused by the
interaction of fields. Such choice of the main approximation leads
to a normally ordered form of the interaction Hamiltonian.
Generation of the fermion mass due to the interaction with exchange
of the scalar boson is investigated. It is demonstrated that, for
zero bare mass, the fermion can acquire mass only if the coupling
constant exceeds the critical value determined by the boson mass. In
this connection, the problem of the neutrino mass is discussed.
\newline%
{\bf Key words}: Yukawa model, fermion, boson, self-consistent
field, perturbation theory, neutrino, mass generation
\end{abstract}
\pacs{11.10.-z, 03.70.+k} %
\maketitle %

{\bf 1.} The theory of interacting Dirac and scalar fields was
suggested by Yukawa for a description of the interaction of nucleons
and pions \cite{Nishidzhima}. In modern particle theory, the
standard model contains the term with the Yukawa interaction which
relates the Higgs scalar field to quarks and leptons, so that the
majority of free parameters of the standard model are the Yukawa
coupling constants \cite{Cheng}. The Yukawa theory can also be used
as a simplified model of quantum electrodynamics. However, since
massive scalar bosons rather than massless vector bosons are
examined, this model contains no difficulties which arise in % 
the process of electromagnetic field quantization \cite{Prokhorov}.

In the present work, a modified perturbation theory for the Yukawa
model is suggested, which is based on choosing the self-consistent
field model as the main approximation. For nonrelativistic
many-particle Fermi and Bose systems, an analogous approach was
developed in works \cite{Poluektov1, Poluektov2}. In
\cite{Poluektov3}, the idea of this approach was realized on the
example of the quantum-mechanical problem of the anharmonic oscillator. %

The Lagrange density function for the Yukawa model has three terms
\begin{equation} \label{E01}
\begin{array}{l}
\displaystyle{%
  L=L_D+L_S+L_I, %
}%
\end{array}
\end{equation}
where
\begin{equation} \label{E02}
\begin{array}{l}
\displaystyle{%
  L_D = -\left( \overline{\psi}\gamma_\mu\partial_\mu\psi+m\overline{\psi}\psi \right) %
}%
\end{array}
\end{equation}
is the Dirac field Lagrangian,
\begin{equation} \label{E03}
\begin{array}{l}
\displaystyle{%
  L_S = -\frac{1}{2}\!\left[ (\partial_\mu\phi)^2+\kappa_0^2\phi^2 \right] %
}%
\end{array}
\end{equation}
is the scalar field Lagrangian, and
\begin{equation} \label{E04}
\begin{array}{l}
\displaystyle{%
  L_I = -g\phi\,\overline{\psi}\psi %
}%
\end{array}
\end{equation}
is the interaction Lagrangian of the scalar and Dirac fields $\phi$
and $\psi$. Here $m$ and $\kappa_0$ are the bare masses of the Dirac
and scalar fields, $g$ is the dimensionless coupling constant,
$\partial_\mu=\partial/\partial x_\mu$, $x_\mu\equiv({\bf x},x_0)$,
and $\gamma_\mu$ are the Dirac matrices. The metric $ab={\bf a}{\bf
b}-a_0b_0$ and the system of units $\hbar=c=1$ are used. Field
operators entering into the initial Lagrangian (\ref{E01}) are
written down in the Heisenberg representation and obey the standard
commutation conditions at coinciding times.\vspace{1mm}

{\bf 2.} The Yukawa model, like the majority of field theories, can
be studied within the framework of perturbation theory. As a rule,
the free field Lagrangian $L_0=L_D+L_S$ is used as the main
approximation, and the interaction (\ref{E04}) is considered as a
perturbation. The validity of this approach is substantiated the
more rigorously the smaller is the coupling constant. Meanwhile, in
the models with the Yukawa interaction, this interaction can be
strong so that the perturbation theory in its conventional form
proves to be, strictly speaking, inapplicable. %
However, a ``natural'' decomposition of the Lagrangian into the main
part and the perturbation is not unique and necessary. %
Indeed, instead of some decomposition $L=L_1+L_2$, we can use as a
basis another one $L=L_1'+L_2'$, where $L_1'=L_1+\Delta L$,
$L_2'=L_2-\Delta L$, and $\Delta L$ is some operator addition. By
fixing the way of constructing the Lagrangian which describes the
one-particle states, we thereby in fact define the notion of
``noninteracting particle'' within the framework of the nonlinear
theory.

We now consider the influence of the perturbation (\ref{E04}) on the
free Dirac and scalar fields. The structure of the perturbation
(\ref{E04}) is such that it has the form of the mass term (with the
operator coefficient) involved in the Lagrangian $L_D$. Therefore,
we can assume that this interaction will lead to a change in the
particle mass described by the field $\psi$. To consider this
effect, we introduce the Lagrangian
\begin{equation} \label{E05}
\begin{array}{l}
\displaystyle{%
  L_D' = -\left( \overline{\psi}\gamma_\mu\partial_\mu\psi+M\overline{\psi}\psi \right), %
}%
\end{array}
\end{equation}
where $M$ is the mass different from the bare particle mass $m$ of
the Lagrangian (\ref{E02}). We note also that the interaction
(\ref{E04}) is linear in the scalar field and thus breaks the
symmetry of the Lagrangian $L_S$ with respect to a change in the
scalar field sign: $\phi\rightarrow -\phi$. To consider this effect,
we introduce the Lagrangian
\begin{equation} \label{E06}
\begin{array}{l}
\displaystyle{%
  L_S' = -\frac{1}{2}\!\left[ (\partial_\mu\phi)^2+\kappa^2\phi^2 \right]-b\phi, %
}%
\end{array}
\end{equation}
where $b$ is a real coefficient. The mass $\kappa$ has been
introduced instead of $\kappa_0$ in Eq. (\ref{E06}), assuming for
generality that the interaction can change the scalar particle mass
as well. However, as shown below, the structure of the interaction
(\ref{E04}) is such that this change does not actually occur. Thus,
the Lagrangians (\ref{E05}) and (\ref{E06}) consider to a certain
degree the effects caused by the interaction of fields. As the main
approximation, we choose the Lagrangian which is a sum of the
Lagrangians (\ref{E05}) and (\ref{E07}):
\begin{equation} \label{E07}
\begin{array}{l}
\displaystyle{%
  L_0 = -\!\left( \overline{\psi}\gamma_\mu\partial_\mu\psi+\!M\overline{\psi}\psi \right)
  \!-\!\frac{1}{2}\!\left[ (\partial_\mu\phi)^2\!+\!\kappa^2\phi^2 \right]\!-\!b\phi-\!V, %
}%
\end{array}
\end{equation}
where $V$ is the $c$\,-number constant which is important for
constructing the perturbation theory in this approach. With this
choice of the main approximation, the perturbation Lagrangian is
evidently $L_C=L-L_0$, so that
\begin{equation} \label{E08}
\begin{array}{l}
\displaystyle{%
  L_C = -g\phi\,\overline{\psi}\psi+(M-m)\overline{\psi}\psi-\!\frac{1}{2}\!\left(\kappa_0^2-\kappa^2\right)\phi^2+b\phi+\!V. %
}%
\end{array}
\end{equation}
Thus, total Lagrangian (\ref{E01}) is a sum of terms $L=L_0+L_C$. %
In fact, the expression $\Delta L=-M\overline{\psi}\psi-\kappa^2\phi^2/2-b\phi-V$ %
has been added and subtracted, so that the total Lagrangian
(\ref{E01}) remained unchanged, and thus until now no approximations
have been used. Now arbitrary parameters $M, \kappa, b$ and $V$ have
appeared in the Lagrangian, which have to be found from additional considerations. %

The total Hamiltonian $H=H_0+H_C$ can be represented similarly:
\begin{equation} \label{E09}
\begin{array}{l}
\displaystyle{%
  H_0 = \int\!d{\bf  x}\bigg[\frac{1}{2}\pi^2+\frac{1}{2}(\nabla\phi)^2+\frac{1}{2}\kappa^2\phi^2 + b\phi\,\,+ %
}\vspace{1mm}\\ %
\displaystyle{\hspace{35mm}%
  +\,\,\overline{\psi}(\bm{\gamma}\nabla)\psi+M\overline{\psi}\psi+V\bigg],  %
}%
\end{array}
\end{equation}
\begin{equation} \label{E10}
\begin{array}{l}
\displaystyle{%
  H_C = \int\!d{\bf  x}\bigg[g\phi\,\overline{\psi}\psi - b\phi + \frac{1}{2}\!\left(\kappa_0^2-\kappa^2\right)\!\phi^2+%
}\vspace{1mm}\\ %
\displaystyle{\hspace{40mm}%
  +(m-M)\overline{\psi}\psi-V\bigg].  %
}%
\end{array}
\end{equation}
Since the total Hamiltonian is expressed in the same way both
through Heisenberg and Schr\"{o}dinger operators, then $H_0$ and
$H_C$ in Eqs. (\ref{E09}) and (\ref{E10}) can also be expressed
through field operators in these representations. \vspace{0mm}

{\bf 3.} Let us consider in more detail the main approximation
determined by Hamiltonian (\ref{E09}), expressed through the
Schr\"{o}dinger operators. This approximation describes
``noninteracting'' particles in our understanding. We now determine
the operators in the interaction representation:
\begin{equation} \label{E11}
\begin{array}{l}
\displaystyle{%
  \hat{\phi}({\bf x},t)=e^{iH_0t}\hat{\phi}({\bf x},0)e^{-iH_0t},%
}\vspace{2mm}\\ %
\displaystyle{%
  \hat{\pi}({\bf x},t)=e^{iH_0t}\hat{\pi}({\bf x},0)e^{-iH_0t},%
}\vspace{2mm}\\ %
\displaystyle{%
  \hat{\psi}({\bf x},t)=e^{iH_0t}\hat{\psi}({\bf x},0)e^{-iH_0t}.%
}
\end{array}
\end{equation}

The simultaneous commutation relationships for operators (\ref{E11})
have the same form as for the Heisenberg operators. Obviously, $H_0$
is expressed through operators (\ref{E11}) in the same way as
through the Schr\"{o}dinger operators. In the following it is
convenient to proceed to a new ``shifted'' scalar field $\varphi$ by
representing the initial field in the form $\phi=\varphi+\chi$,
where $\chi$ is the $c$\,-number. This substitution allows us to
eliminate the term linear in field $\phi$ from the Hamiltonian
(\ref{E09}) by the proper choice of $\chi$. As a result, the
Hamiltonian of the main approximation in the interaction
representation assumes the form
\begin{equation} \label{E12}
\begin{array}{l}
\displaystyle{%
  \hat{H}_0 = \int\!d{\bf  x}\bigg[\frac{1}{2}\hat{\pi}^2+\frac{1}{2}(\nabla\hat{\varphi})^2+\frac{1}{2}\kappa^2\hat{\varphi}^2\, + %
}\vspace{1mm}\\ %
\displaystyle{\hspace{12mm}%
  +\,\,\overline{\hat{\psi}}(\bm{\gamma}\nabla)\hat{\psi}+M\overline{\hat{\psi}}\hat{\psi}+ \frac{1}{2}\kappa^2\chi^2+b\chi+V\bigg].  %
}%
\end{array}
\end{equation}
The condition
\begin{equation} \label{E13}
\begin{array}{l}
\displaystyle{%
  b+\kappa^2\chi=0
}%
\end{array}
\end{equation}
must be fulfilled that provides the elimination of the linear in the
field $\phi$ term in Eq.\,(\ref{E12}). If $|0\rangle$ is the vacuum
state vector of the system described by the Hamiltonian (\ref{E12}),
then since only terms quadratic in the field $\hat{\varphi}$ enter
into Eq.\,(\ref{E12}), the field averaged over the vacuum state is
equal to zero: $\big\langle\hat{\varphi}\big\rangle_0\equiv\big\langle0|\hat{\varphi}|0\big\rangle=0$. %
From here it follows that the parameter $\chi$ has the meaning of
the vacuum average of the initial scalar field: $\chi=\big\langle\hat{\phi}\,\big\rangle_0$. %
The relations
\begin{equation} \label{E14}
\begin{array}{l}
\displaystyle{%
  \big\langle\hat{\phi}\,^2\big\rangle_0=\big\langle\hat{\varphi}^2\big\rangle_0 + \chi^2, \quad%
  \big\langle\hat{\phi}\,\overline{\hat{\psi}}\hat{\psi}\big\rangle_0=\chi\big\langle\overline{\hat{\psi}}\hat{\psi}\big\rangle_0. %
}%
\end{array}
\end{equation}
can also be established easily.

Until now the parameters $M, \kappa, b$ and $V$ were not in any way
fixed. A method for determining these parameters should be
specified. Having indicated this method, we in fact determine the
method of constructing one-particle states within the framework of
our approach. The more successful the choice of these parameters,
the more efficient the perturbation theory will prove to be. Let us
postulate that parameters $M, \kappa, b$ and $V$ are determined from
the requirement that the approximating Hamiltonian $H_0$ is most
close in a certain sense to the exact Hamiltonian of the system
under study. For this we require that the vacuum average of the
difference between the exact and approximating Hamiltonians, that
equals to the vacuum average of the interaction Hamiltonian %
$I\equiv \big\langle H_C\big\rangle_0$, is minimal. Thus, to
determine the unknown parameters we use the following conditions:
\begin{equation} \label{E15}
\begin{array}{l}
\displaystyle{%
  \frac{\partial I}{\partial \big\langle\hat{\phi\,}\big\rangle_0}=0,\quad %
  \frac{\partial I}{\partial \big\langle\hat{\phi}\,^2\big\rangle_0}=0,\quad %
  \frac{\partial I}{\partial \big\langle\overline{\hat{\psi}}\hat{\psi}\big\rangle_0}=0. %
}%
\end{array}
\end{equation}
Moreover, we require that the average over the vacuum state of the
interaction Lagrangian be equal to zero: $\langle0|H_C|0\rangle=0$. %
The last requirement is caused by the fact that, since $H_C$
describes the particle interaction, then, naturally, the energy of
this interaction in the vacuum state must be equal to zero. The
account for the non-operator term $V$ provides the fulfillment of
this condition. The listed requirements lead to the following relations %
\begin{equation} \label{E16}
\begin{array}{l}
\displaystyle{%
  b=g\big\langle\overline{\hat{\psi}}\hat{\psi}\big\rangle_0,\qquad\quad \kappa_0=\kappa,\, %
}\vspace{2mm}\\ %
\displaystyle{\hspace{0mm}%
  M=m+g\big\langle\hat{\phi\,}\big\rangle_0,\quad %
  V=-(M-m)\big\langle\overline{\hat{\psi}}\hat{\psi}\big\rangle_0. %  %
}%
\end{array}
\end{equation}
As we see, the interaction of the form (\ref{E04}), as it was
expected, does not lead to a change in the scalar particle mass.
Note that this circumstance is caused by the structure of the
interaction (\ref{E04}). If the interaction of fields had, for
example, the form $-g\phi^2\overline{\psi}\psi$, or %
the scalar field Lagrangian comprised the self-action of the form $\phi^4$, %
the scalar particle mass would differ from the bare mass $\kappa_0$.
From the first formula of (\ref{E16}), accounting for (\ref{E13}), we derive%
\begin{equation} \label{E17}
\begin{array}{l}
\displaystyle{%
  \kappa^2\chi+g\big\langle\overline{\hat{\psi}}\hat{\psi}\big\rangle_0=0.%
}%
\end{array}
\end{equation}
Considering Eq.\,(\ref{E17}), from (\ref{E16}) we obtain the relation %
\begin{equation} \label{E18}
\begin{array}{l}
\displaystyle{%
  M=m-\big(g^2\big/\kappa^2\big)\big\langle\overline{\hat{\psi}}\hat{\psi}\big\rangle_0. %
}%
\end{array}
\end{equation}

{\bf 4.} The field operator in the interaction representation, the
same as for free particles, can be represented as a Fourier integral %
\begin{equation} \label{E19}
\begin{array}{l}
\displaystyle{%
  \hat{\psi}(x)=(2\pi)^{-3/2}\!\sum_{r=\pm 1}\int\!d{\bf p}\big(\!M\big/p_0\big)^{1/2}\times} %
\vspace{1mm}\\ %
\displaystyle{\hspace{13mm}%
  \times\Big[u^r(p)c_r(p)e^{ipx}+u^r(-p)d_r^+(p)e^{-ipx}\Big],  %
}%
\end{array}
\end{equation}
where $c_r(p)$ and $d_r^+(p)$ are the operators of annihilation of
particles and creation of antiparticles with helicity $r$,
$p_0=\sqrt{M^2+{\bf p}^2}$, and $p=({\bf p},p_0)$. %
Using the representation (\ref{E19}), we derive the vacuum average
\begin{equation} \label{E20}
\begin{array}{l}
\displaystyle{%
  \Big\langle\overline{\hat{\psi}}(x)\hat{\psi}(x)\Big\rangle_0 = %
  -\frac{M}{4\pi^3}\int\!\frac{d{\bf p}}{\sqrt{M^2+{\bf p}^2}} = %
}\vspace{2mm}\\ %
\displaystyle{\hspace{21mm}%
  = \lim_{\varepsilon\rightarrow\,+0}\frac{iM}{4\pi^4}\int\!\frac{dp}{p^2+M^2-i\varepsilon}.  %
}%
\end{array}
\end{equation}
As is known \cite{Nambu}, the integral in Eq.\,(\ref{E20}) diverges
at large momenta. As a consequence, there appears a need to
introduce the cutoff parameter $\Lambda$. The cutoff parameter have
been introduced in early works \cite{Nambu,VL} where the methods of
superconductivity theory were for the first time used to construct
models of elementary particles. Note that we have to cut off the
integral at large momenta also in the self-consistency equation in
the superconductivity theory. Here, however, natural characteristic
scales are present -- the average interparticle distance and the
radius of action of the interparticle potential. Let us introduce
the cutoff parameter in a relativistically invariant way. When
calculating the integral in (\ref{E20}), the Wick rotation of the
integration contour is performed and the substitution $p_0=ip_4$ is
used \cite{Cheng}. As a result, we obtain
\begin{equation} \label{E21}
\begin{array}{l}
\displaystyle{%
  \Big\langle\overline{\hat{\psi}}(x)\hat{\psi}(x)\Big\rangle_0 = %
  -\big(2M\Lambda^2\big/\pi^2\big)f\big(\tilde{M}^2\big), %
}
\end{array}
\end{equation}
where $\tilde{M}^2=M^2\big/\Lambda^2$. The function
\begin{equation} \label{E22}
\begin{array}{l}
\displaystyle{%
  f\big(\tilde{M}^2\big)=1-\tilde{M}^2\ln\!\big(1+\tilde{M}^{-2}\big) %
}
\end{array}
\end{equation}
is equal to unity for $\tilde{M}^2=0$ and monotonically decreases
approaching zero for $\tilde{M}^2\rightarrow\infty$. Substituting
the expression for the vacuum average (\ref{E21}) into (\ref{E18}),
we obtain the self-consistency equation that determines the fermion mass: %
\begin{equation} \label{E23}
\begin{array}{l}
\displaystyle{%
  1-\tilde{m}\big/\tilde{M}=G f\big(\tilde{M}^2\big), %
}
\end{array}
\end{equation}
where $G=2g^2\big/\pi^2\tilde{\!\kappa}^2$, $\tilde{m}=m/\Lambda$,
$\tilde{\!\kappa}=\kappa/\Lambda$. As we see, the fermion mass is
independent of the sign of the Yukawa coupling constant and is
determined also by the scalar particle mass. It is convenient to
analyze Eq.\,(\ref{E23}) graphically (Fig.\,1). The solution is the
coordinate of the intersection point of the curves
$y_1(\tilde{M})=1-\tilde{m}\big/\tilde{M}$ and $y_2(\tilde{M})=G
f\big(\tilde{M}^2\big)$. For nonzero bare mass $m\neq 0$,
Eq.\,(\ref{E23}) has a solution at an arbitrary value of the
coupling constant. For $G\ll 1$, a small addition to the bare mass
arises: $\tilde{M}\approx\tilde{m}+\Delta\tilde{M}$, where
$\Delta\tilde{M}=G\tilde{m}f\big(\tilde{m}^2\big)$. In the opposite
case $G\gg 1$, we have $\tilde{M}^2\approx G/2$. The fermion mass
decreases with increasing the boson mass and for large
$\tilde{\kappa}$ tends to its bare mass, whereas for small
$\tilde{\kappa}$ the fermion mass increases inversely proportional
to $\tilde{\kappa}$ (curve {\it 1} in Fig.\,2). For a fixed boson
mass, the fermion mass increases monotonically with increasing the
coupling constant (curve {\it 1} in Fig.\,3). \vspace{-0mm}
\begin{figure}[h!]
\centering %
  \includegraphics[width =1.18\columnwidth]{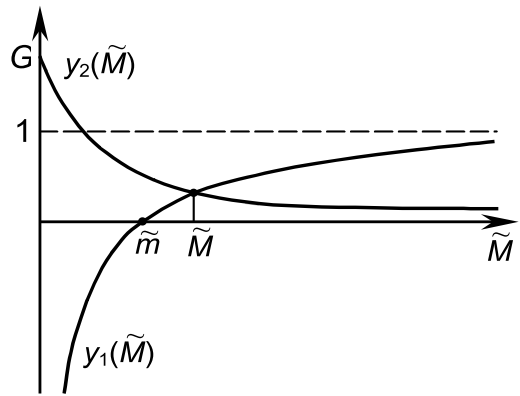} %
  %\vspace{0cm}\scalebox{0.6}[0.6]{\includegraphics[bb = 47 600 366 777]{Fig01.eps}} %
\vspace{-10mm}
\caption{\label{fig01} %
Graphic solution of Eq.\,(23).
}%
\end{figure}

\newpage
{\bf 5.} The case of zero bare mass of fermions should be examined
separately. In this case, the equation determining the fermion mass
assumes the form
\begin{equation} \label{E24}
\begin{array}{l}
\displaystyle{%
  M\big(1-G f\big(\tilde{M}^2\big)\big)=0, %
}
\end{array}
\end{equation}
so that the solution $M=0$ always exists. For $G<1$, this solution
is unique. Thus, if $m=0$, then provided
$g^2<\pi^2\tilde{\!\kappa}^2/2$ the fermion mass cannot arise as a
result of the interaction. For $G>1$, in addition to the solution
$M=0$, Eq.\,(\ref{E24}) has the solution with $\tilde{M}\neq 0$
determined by the equation
\begin{equation} \label{E25}
\begin{array}{l}
\displaystyle{%
  G^{-1}=f\big(\tilde{M}^2\big). %
}
\end{array}
\end{equation}
Dependence of the fermion mass on $\tilde{\!\kappa}$ for $m=0$ is
shown in Fig.\,2 (curve {\it 2}). The mass $\tilde{M}$ decreases
with increasing the scalar particle mass and vanishes for
$\tilde{\!\kappa}=\sqrt{2}\,g/\pi$. For large values of
$\tilde{\!\kappa}$, the unique solution $M=0$ exists. To find out
which from two solutions existing at small $\tilde{\!\kappa}$
corresponds to the stable state, we have to calculate the vacuum
energy in both states. Obviously, the state with smaller
ground-state energy will be stable. It should be stressed that the
considered mechanism of mass generation for particles interacting
with the scalar field differs from the mechanism caused by
spontaneous symmetry breaking \cite{Cheng}.\vspace{1mm}
\begin{figure}[t!]
\vspace{03mm}
\centering %
  \includegraphics[width =1.25\columnwidth]{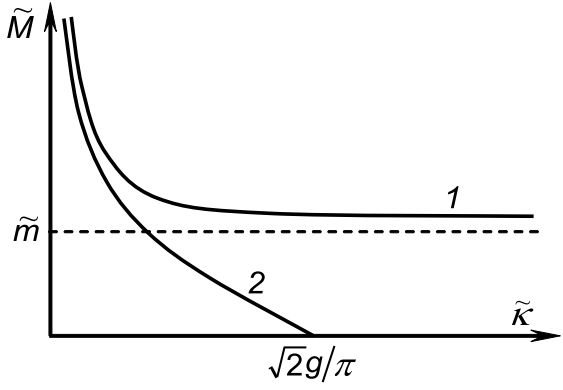} %
\vspace{-11mm}
\caption{\label{fig02} %
Dependence of the fermion mass on the scalar particle mass.
}%
\end{figure}

{\bf 6.} With account of the expansion (\ref{E19}) and the similar
expansion of the scalar field into a Fourier integral
\begin{equation} \label{E26}
\begin{array}{l}
\displaystyle{%
  \hat{\varphi}(x)=(2\pi)^{-3/2}\!\int\!d{\bf q}(2q_0)^{-1/2}\!\big[a(q)e^{iqx}+a^+(q)e^{-iqx}\big], %
} %
\end{array}
\end{equation}
where $a^+(q)$ and $a(q)$ are the operators of creation and
annihilation of the scalar particles, $q_0=\sqrt{\kappa^2+{\bf
q}^2}$, the Hamiltonian (\ref{E12}) can be written in the form
\begin{equation} \label{E27}
\begin{array}{l}
\displaystyle{%
  H_0=\!\sum_{r=\pm 1}\int\!d{\bf p}\,p_0\Big[c_r^+(p)c_r(p)+d_r^+(p)d_r(p)\Big] +} %
\vspace{1mm}\\ %
\displaystyle{\hspace{06mm}%
  +\int\!d{\bf q}\,q_0 a^+\!(q)a(q) + C_M + C_\kappa + V\kappa^2\chi^2\big/2, %
}%
\end{array}
\end{equation}
where
\begin{equation} \label{E28}
\begin{array}{l}
\displaystyle{%
  C_M=-2\int\!d{\bf p}d{\bf p}'\,p_0\,\delta({\bf p}-{\bf p}')\delta({\bf p}-{\bf p}'),} %
\vspace{1mm}\\ %
\displaystyle{\hspace{00mm}%
  C_\kappa=(1/2)\int\!d{\bf q}d{\bf q}'\,q_0\,\delta({\bf q}-{\bf q}')\delta({\bf q}-{\bf q}'). %
}%
\end{array}
\end{equation}
A non-operator part of the Hamiltonian (\ref{E27}) determines the
ground state (vacuum) energy: $E_V=C_M+C_\kappa+V\kappa^2\chi^2\big/2$. %
The first two terms in $E_V$ appeared due to transition to the
normal ordering of operators in $H_0$, and the third term is caused
by the interaction effects. The constants $C_M$, $C_\kappa$ are
obviously infinite. Since the cutoff at large momenta was introduced
above, they can be calculated by means of regularization of the
integral by the same way as in calculation of the integral in the
vacuum average in Eq.\,(\ref{E20}). Considering that
$\delta(0)=V\big/(2\pi)^3$, the constant $C_\kappa$ can be
represented in the form
\begin{equation} \label{E29}
\begin{array}{l}
\displaystyle{%
  C_\kappa=\frac{V}{2(2\pi)^3}\int\!d{\bf q}\sqrt{{\bf q}^2+\kappa^2} = \frac{VJ_\kappa}{2(2\pi)^3}, %
}%
\end{array}
\end{equation}
where $J_\kappa\equiv\int\!d{\bf q}\sqrt{{\bf q}^2+\kappa^2}$.
Having differentiated $J_\kappa$ with respect to $\kappa^2$, we
arrive at the integral (\ref{E22}) obtained above so that
\begin{figure}[t!]
\vspace{3.5mm}
\centering %
  \includegraphics[width =1.25\columnwidth]{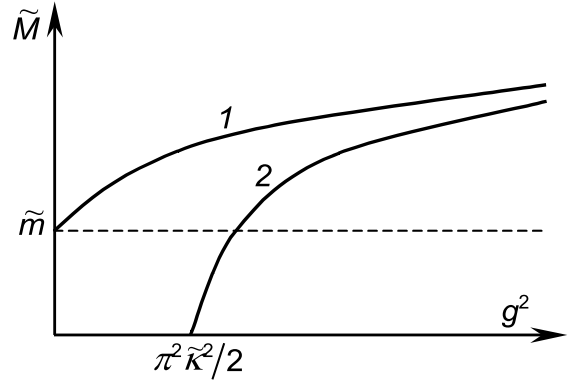} %
\vspace{-12mm}
\caption{\label{fig03} %
Dependence of the fermion mass on the squared coupling constant.
}%
\end{figure}
\begin{equation} \nonumber
\begin{array}{l}
\displaystyle{%
  \frac{d J_\kappa}{d\kappa^2} = \frac{1}{2}\int\!\frac{d{\bf q}}{\sqrt{{\bf q}^2+\kappa^2}} = %
  4\pi\Lambda^2\big[1-\tilde{\!\kappa}^2\ln\!\big(1+\tilde{\!\kappa}^{-2}\big)\big].  %
}%
\end{array}
\end{equation}
Having integrated this relation, we obtain
\begin{equation} \label{E30}
\begin{array}{l}
\displaystyle{\hspace{-2.0mm}%
  C_\kappa=\!V\varepsilon_0\big[\!-\tilde{\!\kappa}^4\ln\!\big(1+\tilde{\!\kappa}^{-2}\big) %
  \!+\ln\!\big(1+\tilde{\!\kappa}^2\big)\! + \tilde{\!\kappa}^2 + c_\kappa' \big], %
}%
\end{array}
\end{equation}
where $c_\kappa'$ is an integration constant independent of the
system parameters, which can be set equal to zero. In this case the
constant $C_\kappa$ is always positive for $\tilde{\!\kappa}^2>0$
and increases monotonically from zero with increasing $\tilde{\!\kappa}^2$. %
The constant $C_M$ can be calculated similarly, which, according to
(\ref{E28}), has the opposite to $C_\kappa$ sign. Finally, the
vacuum state energy can be represented in the form
\begin{equation} \label{E31}
\begin{array}{ll}
\displaystyle{\hspace{0mm}%
  E_V\big/V\varepsilon_0 = \tilde{\!\kappa}^{2}\!f\big(\tilde{\!\kappa}^2\big)+\ln\!\big(1+\tilde{\!\kappa}^2\big)\,- %
}\vspace{3mm}\\ %
\displaystyle{\hspace{4mm}%
  -\,4\big[\tilde{M}^2\!f\big(\tilde{M}^2\big)\!+\ln\!\big(1+\tilde{M}^2\big)\big]\!+8\tilde{M}\big(\tilde{M}-\tilde{m}\big)f\big(\tilde{M}^2\big). %
}%
\end{array}
\end{equation}
Here the function $f(x)$ is determined by the formula (\ref{E22})
and $\varepsilon_0=\Lambda^4\big/8\pi^2$. Formula (\ref{E31})
together with formula (\ref{E23}) determines the vacuum energy
density as a function of the dimensionless parameters $\tilde{m}$,
$\tilde{\kappa}$, $g^2$. As is seen, in the case $m\neq 0$ (curve
{\it 1} in Fig. 4) and for small $\tilde{\!\kappa}^2$ the vacuum
energy is negative and increases with increasing the boson mass. At
a certain value $\tilde{\!\kappa}_0$ this energy vanishes and
becomes positive for $\tilde{\!\kappa}>\tilde{\!\kappa}_0$,
continuing its increase with increasing $\tilde{\!\kappa}$. In the
case of zero bare mass $m=0$ (curve {\it 2} in Fig. 4) and for
$\tilde{\!\kappa}^2>2g^2/\pi^2$ the fermion mass is equal to zero
and the vacuum energy is positive and increases with increasing
$\tilde{\!\kappa}$. For $\tilde{\!\kappa}^2<2g^2/\pi^2$, two
solutions exist. One of them (curve {\it 2a} in Fig. 4) corresponds
to the fermion with zero mass, and the second solution (curve {\it
2b} in Fig. 4) corresponds to the fermion with a finite mass. The
stable solution (corresponding to the minimum of the vacuum energy
density) is that with the nonzero fermion mass. Thus, for a
sufficiently large boson mass, in the case of zero bare mass the
fermions remain massless.

\begin{figure}[t!]
\vspace{0mm}
\centering %
  \includegraphics[width =1.25\columnwidth]{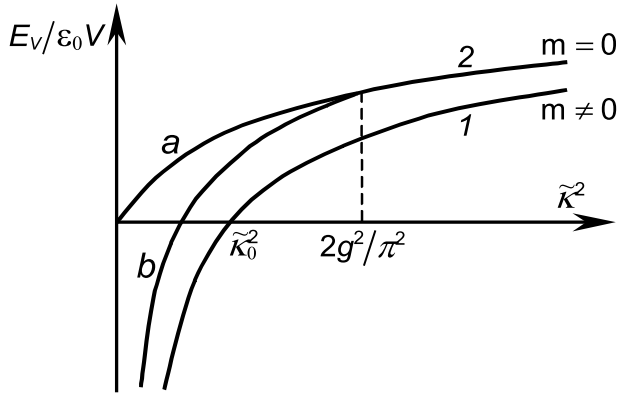} %
\vspace{-07mm}
\caption{\label{fig04} %
Dependence of the vacuum energy density on the squared boson mass.
}%
\end{figure}

Let us consider the dependence of the vacuum energy density on the
coupling constant. For $m\neq 0$ (curve {\it 1} in Fig. 5), the
vacuum energy decreases monotonically with increasing the coupling
constant. In the case $m=0$ (curve {\it 2} in Fig. 5) and for
$g^2<\pi^2\tilde{\!\kappa}^2/2$, when the fermion mass is equal to
zero $(M=0)$, the vacuum energy is independent of the coupling
constant. For $g^2>\pi^2\tilde{\!\kappa}^2/2$, apart from the state
with $M=0$ (curve {\it 2a} in Fig. 5), the state with nonzero
fermion mass and smaller vacuum energy density appears (curve {\it
2b} in Fig. 5). Thus, generation of the fermion mass due to the
interaction with the boson exchange is possible only if the coupling
constant exceeds a certain critical value determined by the boson
mass. At a small coupling constant and zero bare mass, the fermions
remain massless.

We note that in quantum field theory, the infinite vacuum energy
caused by zero oscillations is, as a rule, excluded from
consideration. In our approach to the construction of the
perturbation theory, when the interaction in the self-consistent
field approximation is accounted for already at the stage of
constructing the one-particle states and the equation for the mass
of particles can have several solutions, calculation of the vacuum
energy becomes necessary to have a possibility of choosing a stable
solution. The second reason for calculation of the vacuum energy is
its possible influence on the gravitational effects. Recently, this
problem became especially urgent in connection with the problem of
dark energy in cosmology \cite{Chernin}. \vspace{1mm}

{\bf 7.} With account of the derived relations (\ref{E16}) and
(\ref{E17}), the interaction Hamiltonian (\ref{E10}) assumes the
form of the normally ordered product
\begin{equation} \label{E32}
\begin{array}{l}
\displaystyle{%
  \hat{H}_C=g\!\int\!d{\bf x}\,\hat{\varphi}(x)N\Big[\overline{\hat{\psi}}(x)\hat{\psi}(x)\Big], %
} %
\end{array}
\end{equation}
where the normally ordered product of the field operators at one
point can be presented in the form
$N\Big[\overline{\hat{\psi}}(x)\hat{\psi}(x)\Big]=\overline{\hat{\psi}}(x)\hat{\psi}(x)-\Big\langle\overline{\hat{\psi}}(x)\hat{\psi}(x)\Big\rangle_0$. %
The perturbation theory  with interaction (\ref{E32}) is constructed
in the standard way. The boson propagator has the form
$G_B(q)=(-i)\big(q^2+\kappa^2-i\varepsilon\big)^{-1}$ and the
fermion propagator has the form %
$G_D(p)=i\big(p_\mu\gamma_\mu-M\big)\big/\big(p^2+M^2-\varepsilon\big)$.%
The factor $-ig$ corresponds to the vertex. It is important to
stress that in the proposed formulation of perturbation theory the
fermion mass is not an independent parameter, but determined by the
intrinsic bare mass and also by the boson mass and the coupling
constant (relations (\ref{E23})$\,–-\,$(\ref{E25})). In this
approach, it is also important that the normally ordered form of
presentation of the Lagrange function is not assumed from the
beginning \cite{Bogolyubov}, since the latter means that the vacuum
energy density is initially chosen to be zero. %
The choice of the main approximation in the proposed approach, that
is based on the self-consistent field model, allows to naturally
arrive at the normal form of the interaction Hamiltonian
(\ref{E32}), not assuming that a priory, and gives a possibility to
calculate the vacuum energy (using, certainly, the cutoff parameter $\Lambda$). %
\vspace{1mm}
\begin{figure}[t!]
\vspace{5mm}
\centering %
  \includegraphics[width =1.25\columnwidth]{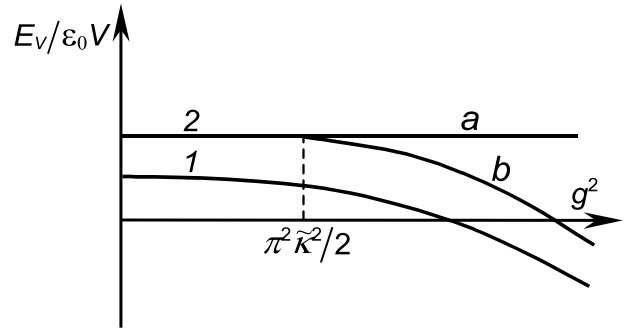} %
\vspace{-6.5mm}
\caption{\label{fig05} %
Dependence of the vacuum energy density on the squared coupling constant. %
}%
\end{figure}

\newpage
{\bf 8.} In conclusion, relying on the results of sections 5 and 6,
we discuss the problem of the neutrino mass. It is considered
\cite{Okun} that, from purely theoretical viewpoint, there are no
any grounds to consider the neutrino masses equal to zero. There is
a widespread belief that a strict local symmetry is necessary in
order for a massless particle to exist, and since such symmetry is
absent in the case of neutrino then zero mass must not exist.

Let us assume that the Dirac field with zero bare mass, considered
above, describes neutrino (neutrino types and other details of
description of the neutrino field are unimportant in this case). As
is well known, the carrier of the weak interaction is the vector
boson \cite{Cheng}. We can assume, however, that change of the
vector boson for the scalar one will not influence considerably on
further conclusions. Then, as follows from the results of section 5,
in order for neutrino to acquire mass the condition
$g^2>\pi^2\kappa^2\big/2\Lambda^2$ between the boson mass and the
coupling constant must be fulfilled. The dimensionless constant for
the case of the weak interaction $g^2\approx 0.4$ can be obtained
from the relation \cite{Cheng} $g^2/8m_W^2=G\big/\sqrt{2}$, where
$m_W\approx 80$\,GeV is the boson mass, $G\approx 10^{-5}\big/m_p^2$
is the Fermi weak coupling constant, $m_p\approx 0.938$\,GeV is the
proton mass. To estimate the parameter $\Lambda$, let us calculate
its value for the case when the equality holds
$g_k^2=\pi^2\kappa^2\big/2\Lambda^2$, setting $\kappa=m_W$. %
It gives $\Lambda\approx 280$\,GeV. This value is close to the
electroweak energy scale, estimated to be in the limits
$\Lambda_{EW}\approx\,300$\,GeV $-$ 1\,TeV \cite{Chernin}. From this
estimate we can conclude that the dimensionless constant of the weak
interaction is close to its critical value. Experimental estimates
of the electron neutrino mass give $M_{\nu e}<35$\,eV \cite{Okun}. %
In dimensionless units this means that $\tilde{M}_{\nu e}<1.3\!\cdot\!10^{-10}$. %
If such small value of the neutrino mass is caused by the weak
interaction, then the squared dimensionless weak interaction
constant exceeds the critical value by a very small quantity
$\big(g^2-g_k^2\big)\big/g_k^2\approx 7.7\!\cdot\!10^{-19}$. This
possibility seems extremely improbable. It should also be noted
that, if the squared dimensionless constant differed from the
critical value only by one percent, then the neutrino mass generated
by the weak interaction would be about 11\,GeV. Such a large mass is
caused by the large value of the electroweak energy scale. It seems
more natural to assume that the weak coupling constant is less than
the critical value (most likely not much) and the strength of the
interaction is insufficient for generation of the neutrino mass.
Thus, the reason of possible zero neutrino mass is not of symmetric
but rather of dynamic character.

%\vspace{20cm}


\begin{thebibliography}{99}
\bibitem{Nishidzhima}
  K.\,Nishijima, Fundamental particles, W.A.\,Benjamin, New York (1964).
  %K.\,Nishidzhima, Fundamental Particles [Russian translation], Mir, Moscow (1965). %
\bibitem{Cheng}
  T.-P.\,Cheng and L.-F.\,Lee, Gauge theory of elementary particle physics, Oxford University Press (1987).  %
 %T.-P.\,Cheng and L.-F.\,Lee, Gauge Theories in Elementary Particle Physics [Russian translation], Mir, Moscow (1987). %
\bibitem{Prokhorov}
  L.V.\,Prokhorov, Quantization of the electromagnetic field, Sov. Phys. Usp. \textbf{31}, $151 - 162$ (1988). % Usp. Fiz. Nauk, \textbf{154}, 299 (1988).
\bibitem{Poluektov1}
  Yu.M.\,Poluektov, On the quantum-field description of many-particle Fermi systems with spontaneously broken symmetry, %
  Ukr. J. Phys. \textbf{50}\,(11), $1303 - 1316$ (2005).
\bibitem{Poluektov2}
  Yu.M.\,Poluektov, On the quantum-field description of many-particle Bose systems with spontaneously broken symmetry, %
  Ukr. J. Phys. \textbf{52}\,(6), $579 - 595$ (2007).
\bibitem{Poluektov3}
  Yu.M.\,Poluektov, Modified perturbation theory of an anharmonic oscillator, %
  Russ. Phys. J., \textbf{47}\,(6), $656 - 663$ (2004).
\bibitem{Nambu}
  Y.\,Nambu and G.\,Jona-Lasinio, Dynamical model of elementary particles based on an analogy with superconductivity, %
  Phys. Rev. \textbf{122}, $345 - 358$ (1961).
\bibitem{VL}
  V.G.\,Vaks and A.I.\,Larkin, On the application of the methods of
  superconductivity theory to the problem of the masses of elementary particles, %
  Sov. Phys. JETP \textbf{13}, $192$ (1961). %Zh. Eksp. Teor. Fiz., \textbf{40}, 282 (1961).
\bibitem{Chernin}
  A.D.\,Chernin, Dark energy and universal antigravitation,
  Phys. Usp. \textbf{51} $253 - 282$ (2008). %Usp. Fiz. Nauk, \textbf{178}, 267 (2008).
\bibitem{Bogolyubov}
  N.N.\,Bogoliubov, D.V.\,Shirkov, Introduction to the theory of quantized fields, Wiley$\&$Sons (1980). %
 %N.N.\,Bogolyubov, D.V.\,Shirkov, Introduction to quantum field theory [in Russian], Nauka,  Moscow (1973). %
\bibitem{Okun}
  L.B.\,Okun, Elementary particle physics [in Russian], Nauka, Moscow (1988). %
\end{thebibliography}
\end{document}